# Ground-based search for the brightest transiting planets with the Multi-site All-Sky CAmeRA - MASCARA

Ignas A. G. Snellen[a], Remko Stuik[a], Ramon Navarro[b], Felix Bettonvil[b], Matthew Kenworthy[a], Ernst de Mooij[c], Gilles Otten[a], Rik ter Horst[b] & Rudolf le Poole[a]

[a]Leiden Observatory, Leiden University, Postbus 9513, 2300 RA, Leiden, The Netherlands
[b]NOVA Optical and Infrared Instrumentation Division at ASTRON, PO Box 2, 7990 AA, Dwingeloo, The Netherlands
[c]Department of Astronomy and Astrophysics, University of Toronto, 50 St. George Street, Toronto, ON M5S 3H4, Canada

## ABSTRACT

The Multi-site All-sky CAmeRA MASCARA is an instrument concept consisting of several stations across the globe, with each station containing a battery of low-cost cameras to monitor the near-entire sky at each location. Once all stations have been installed, MASCARA will be able to provide a nearly 24-hr coverage of the complete dark sky, down to magnitude 8, at sub-minute cadence. Its purpose is to find the brightest transiting exoplanet systems, expected in the V=4-8 magnitude range - currently not probed by space- or ground-based surveys. The bright/nearby transiting planet systems, which MASCARA will discover, will be the key targets for detailed planet atmosphere observations. We present studies on the initial design of a MASCARA station, including the camera housing, domes, and computer equipment, and on the photometric stability of low-cost cameras showing that a precision of 0.3-1% per hour can be readily achieved. We plan to roll out the first MASCARA station before the end of 2013. A 5-station MASCARA can within two years discover up to a dozen of the brightest transiting planet systems in the sky.

## 1. INTRODUCTION

The CoRoT[1] and Kepler[2] space missions are currently revolutionizing exoplanet research. Both missions discovered the first transiting rocky planets (CoRoT-7b[3] and Kepler-10b[4]), and are extending the known transiting planet population to much wider orbits[5]. In particular Kepler is uncovering a zoo of amazing systems. One example is the planet system around Kepler-11, with six transiting planets, of which five closer to their parent star than Mercury in our own solar system – all five with masses less than that of Neptune[6]. Demography based on the list of >2000 Kepler exoplanet candidates suggests that there should be at least tens of billions of planets in the Milky Way.

While CoRoT and Kepler are rapidly revealing the main characteristics of the exoplanet population in our Milky Way, they are not designed to find and study particularly nearby systems. Even with the largest camera ever launched in space (consisting of 95 million pixels), Kepler has 'only' a $115^2$ degree field of view - about 0.25% of the total sky. This has great consequences for the confirmation of exoplanet candidates found by Kepler, and for the further study of their atmospheres. The apparent brightnesses of Kepler-targets are typically of $m_V>11$ magnitude, meaning that, except for those particular planet systems in which planet-planet interactions reveal the planet masses, most of the Kepler objects will remain candidates, allowing only studies of a statistical nature. The fact that Kepler (and CoRoT) only target relatively distant stars has even a stronger effect on atmosphere studies. Even with the largest telescopes on Earth and in space, only very rudimentary atmospheric studies can possibly be performed on Kepler or CoRoT targets.

What the Kepler and CoRoT missions (and radial velocity surveys) *do* teach us is that there should exist a rich and versatile population of transiting planets at bright optical magnitudes. Based on the volume of space targeted by Kepler, and the fact that Kepler probes only 1/400[th] of the sky, an all-sky survey has the potential to find planets ~50 times brighter than Kepler, corresponding to V~5-7 for the brightest transiting Jupiters to rocky planets. The same conclusions can be drawn from the current radial velocity surveys, which find that about 25-30% of nearby solar-type stars host a hot or warm super-Earth/Neptune mass planet[7,8]. With transit probabilities in the range 1-10%, this corresponds to 1 in 100 solar-type stars having a transiting planet, of which the brightest should have an apparent magnitude of V~5-6. Indeed, in 2011, the first member of the naked eye population of transiting planets, *55 Cnc e*, was discovered using optical and near-infrared photometry from space on a known radial velocity planet, using the Canadian MOST satellite and NASA's Spitzer space telescope[9,10].

Currently, transiting planets at these very bright magnitudes can potentially only be found through pointed photometric observations of known radial velocity planets. Up to now, all transiting planet systems with V<8 have been found in this way. However these surveys are typically limited to solar-type stars, and often to the even smaller fraction of stars exhibiting low levels of stellar jitter. The brightest dedicated transit surveys, such as SuperWASP, HATNET, and TrES[11-13], only monitor stars with $m_V$~8 and fainter due to detector saturation limitations.

## 2. THE MASCARA CONCEPT

The Multi-site All-Sky CAmeRA MASCARA will consist of ~5-6 stations across the globe (Figure 1), monitoring the near-entire sky at each location, resulting in the quasi-continuous monitoring of all stars in the V=4-8 magnitude range - the expected sweet-spot of the brightest transiting planet systems. The multi-station approach is required to optimize the time window-function of the observations. From one particular location on Earth a star can only be monitored for typically ~8 hours (30%) per day, while this can be increase up to 100% with several stations spread in longitude. In addition, stations are needed in both the northern and southern hemisphere to obtain all-sky coverage. Each station will consist of multiple combinations of wide-field lenses plus detectors which cover the entire visible sky above a certain airmass. The camera-system will be fixed and not track the sky, and make use of short exposures (a few seconds) to avoid stars producing trails on the detectors. Although this puts significant strains on the required data-analysis pipeline, it enormously simplifies the design and execution of MASCARA. The system will be designed in such way that at the faint end, the photometric precision is equal or better than 1% in one hour, such that the transits of a Jupiter-size planet in front of a solar type star can be detected individually. The photometric precision at the faint magnitude end is set by the (moon dependent) sky background, due to the large solid angle per pixel. At the bright end, the system will produce precisions of 0.1-0.3% per hour, similar to that of other ground-based surveys.

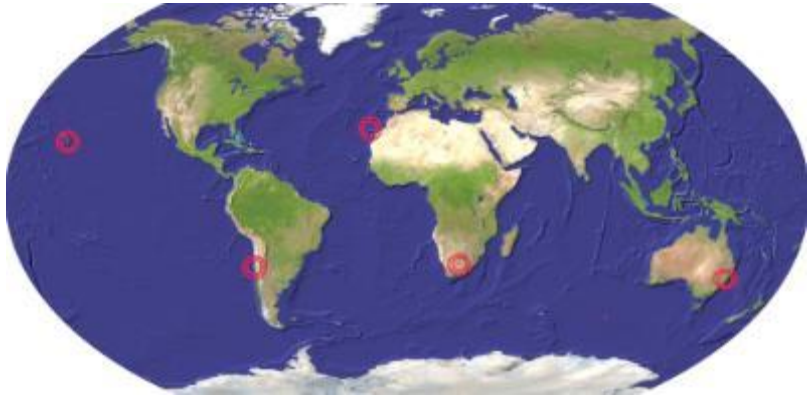

Figure 1. The five locations where we intend to place the MASCARA stations. They are chosen at well-known observatories spread out in longitude and over the northern and southern hemisphere for an optimal time window function.

The stellar light curves will be produced using relative photometry, for which at each moment in time the >1000 visible stars will be used to map the atmospheric absorption as function of azimuth and airmass, and to detect and flag cirrus or heavier cloud cover. In this way MASCARA can also be used as a high-precision real-time extinction monitor for the host observatories. We envisage that in the case the moon is present within the field of one of the cameras, the data of at least that particular camera will be unusable due to saturation and internal reflection issues. The transits of Jupiter-size planets around solar-type stars will be detected individually, meaning that the true detection of a planet is set by the time it takes to observe three or more transits to determine and confirm the orbital period. A planet system with a larger size ratio will require longer observations. For example, 55 cnc[9,10] will be monitored at a precision of 0.2% per hour. The super-Earth orbiting this star produces ~2 hr-long transits with a depth of 0.04%, every 18 hours. This means that a fully operational MASCARA, after one (four) years, would produce a ~3 (~7) sigma detection of its transit.

Figure 2 shows the stellar population targeted by MASCARA, and the expected yield of transiting hot/warm Jupiters, Neptunes, and super-Earths as function of apparent magnitude, simulated using input from the radial velocity surveys. Assuming a 5-site MASCARA as defined in this proposal running for 4 years, the project will yield up to half a dozen of each planet class.

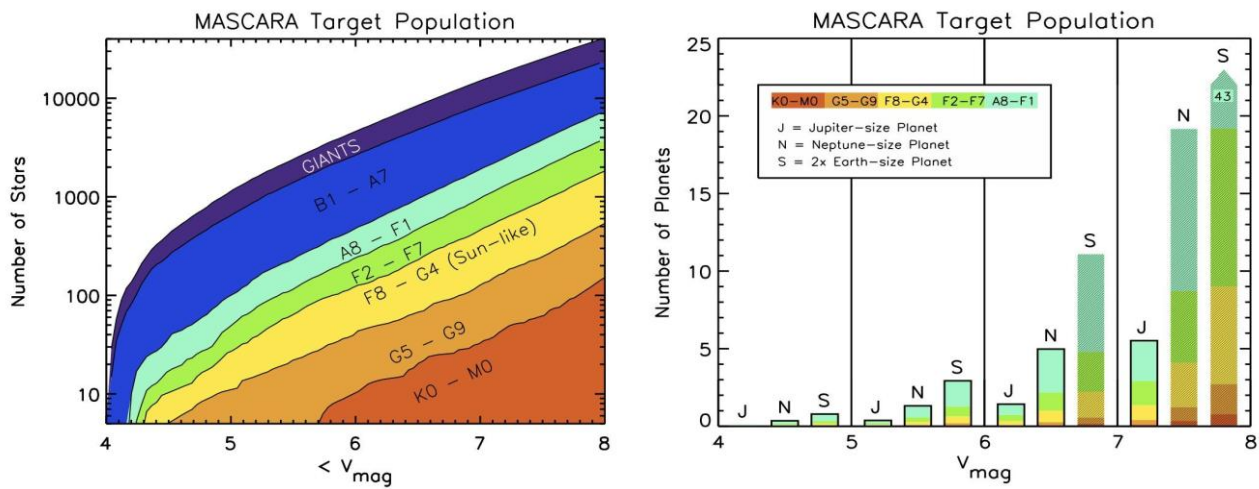

Figure 2. (left panel) The cumulative stellar population as targeted by MASCARA in the magnitude range V=4-8, subdivided in spectral type. (right panel) The expected transiting planet (Jupiter, Neptune, and super-Earths) population as function of magnitude and stellar type (as indicated by the colours). The darkened histograms will be out of reach of a 4-year MASCARA. It indicated that the project will yield up to half a dozen of each planet class.

## 3. MASCARA TEST PHOTOMETRIC OBSERVATIONS

We tested whether high precision photometric time series could be obtained using off-the-shelf consumer cameras. With this aim, a Pentax K10D Digital Single-Lens Reflex camera was used, which utilizes a 3900x2600 pixel Sony ICX-493 AQA CCD, a pixel size of 6 micron, and a 12-bit A/D converter. It was outfitted with a 35 mm f/2.8 (pupil 1.2 cm$^2$) lens, yielding a pixel scale of 36 arcseconds squared and a field of view of ~1000 square degrees. The camera, positioned on an ordinary tripod (without tracking motor) was used during a bright night on the Roque de los Muchachos Observatory on La Palma, and a dark night on mount Paranal in Chile. Each hour, the camera was changed in direction to keep a large fraction of the same stars in the field of view, to create overall time series of 4 hours (La Palma) and 3 hours (Paranal). To prevent the stars from trailing on the CCD, exposures of 10 seconds were used, yielding 970 and 885 images on La Palma and Paranal respectively. In addition a dozen flat-field images were taken during twilight. The La Palma observations were taken during a transit of the famous hot Jupiter[14] HD 189733b (V=7.65), which exhibits transits with a depth of almost 3%.

Since the CCD is equipped with a Bayer colour array, a "white" image is created by summing the flux in the four separate colour-pixels (blue, green1, green2, and red) together, resulting in a 1900x1300 pixel array, with a pixel scale of 72 arcseconds. We performed aperture photometry on the stars in the field using the APER procedure in IDL, with an aperture radius of 3 pixels. Subsequently, relative photometry on a particular star was achieved for each exposure by dividing each photometric data point through the average of that of an ensemble of nearby stars, after which the light curve was binned over 75 measurements to achieve a cadence of ~12.5 minutes.

The resulting light curve for the transiting exoplanet system HD 189733 is shown in Figure 3 (left panel). Superimposed (the smooth curve) are data taken at the same time with the Wide Field Camera on the Isaac Newton Telescope through the RGO U-filter. We detect the transit with the PENTAX camera at about a 3 sigma level. Extrapolating this to a multi-site MASCARA system in full operation, we will catch ~50 transits a year from this planet – accumulating to a 20 sigma detection, showing that we should easily find hot Jupiter systems, even at the faint magnitude limit of MASCARA. The right panel of Figure 3 shows the photometric uncertainty achieved with these test observations as function of apparent magnitude, showing the ~1%/hour limit at V=8.

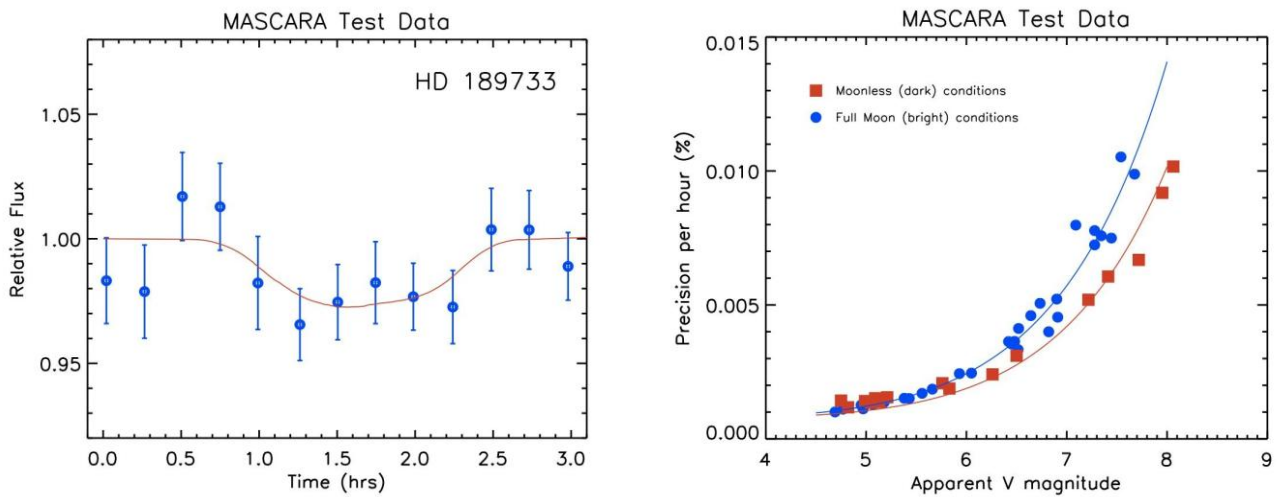

Figure 3: (left panel) The lightcurve of the transiting exoplanet system HD 189733 from test observations with an of-the-shelve consumer camera, showing a 3-sigma detection of its transit. It implies that a fully-operational MASCARA, which will accumulate typically 50 transits a year for such a planet, will easily detect hot Jupiters – even at its faint magnitude limit. (right panel): The observed scatter in the light curve as function of apparent magnitude for both the moonless and moonlit night. The solid lines indicate the best fit to the data.

## 4. PRELIMINARY DESIGN

To achieve the scientific goals of MASCARA as outlined above, a camera system is required at each station with continuous near all-sky coverage, which can observe at high cadence to prevent stars from trailing, with a sufficient dynamic range such that stars between V=4-8 can be imaged without saturation. Furthermore, such system should be highly reliable, flexible, and of low cost. In principle, a combination of consumer camera and lens (e.g. Canon 5D mark II + Canon lens) is ideal. However, in addition to inflexibility in accessing raw data for such a camera, the main problem is shutter life time – which is typically $~10^5$ exposures for this type of consumer camera. Taking into account that a camera in MASCARA will take one exposure per ~3 seconds, it will typically break down after only 10 nights of observing. With five stations, of four or five cameras each, this is unacceptable for MASCARA.

We therefore looked into the option of using a combination of cooled amateur astronomy cameras with off-the-shelf camera lenses (Canon) for which the interfaces are commercially available. Several trade-offs have to be kept in mind. The camera should have a minimum exposure time of 3 seconds or less, and be able to transfer the data on the same time scale. It turns out that a system with many small-size narrow-angle system is less cost effective than a few large-size wide-angle cameras, while a single fish-eye is more sensitive to the moon and system failure (standard 24x36 mm array plus 20 or 24 mm lens is optimal). The pixel size is less important since the sky background and lens resolution dominate over read noise and pixel pitch (i.e. a camera with 29 million of 5.5 micron-size pixels performs similarly to one with 11 million 9 micron-size pixels). Since the full-well capacity approximately scales with pixel size, it does not influence the level of saturation either, for the star-size is set by the lens resolution. To avoid rapid breakdown of the shutters, the system should be able to perform shutterless, which leads to frame-transfer CCDs. In addition, monochrome cameras are typically an order of magnitude more sensitive than those with coloured pixels, although the latter would allow for 3-band photometry of the stars.

After careful trade-off of the pros and cons we narrowed down our potential choice to three cameras: 1) the Atik 11000-M, modified to allow high readout speed, 2) the FLI Micronline ML29050M, and 3) the Apogee A16000. We currently test the Atik 11000-M camera for its real-world performance. An additional advantage of a lower pixel count is the data rate, which is a factor 2.5 less than in the case of a 29 mega-pixel camera. With this camera we expect to reach a relative photometric precision of 1% per hour for a V=9 star, and avoid saturation for all stars fainter than $m_V$=2-3. MASCARA is expected to create ~1 Tbyte of data per station per night, or 5 Tbyte per night for the full MASCARA array. This is too much to transport directly to a central archive and too much to store locally without significant investments in data storage. Currently it is foreseen to have a local storage of ~1 week of the full data of MASCARA and only transmit the data of all monitored stars (~100,000) and transient events to a central archive. This can reduce the data rate by at least a factor of 10, depending on the region of interest that will be stored for each monitored star.

We are also investigating the different trade-offs for the of the MASCARA housing. Its main purpose is to protect the MASCARA cameras from wind, cold and precipitation. Pictures of different design options are shown in Figure 4. It requires an on-board weather station (or a befriended station nearby) to be able to open/close depending on the atmospheric circumstances. It will contain the cameras plus lenses, all pointed to fixed positions in the sky, the camera control-computers, and the control hardware of the open/closure of the housing. Particular thought is going into the temperature control of the cameras, computers and housing, to prevent the build-up of moisture, freezing, and/or overheating of the systems.

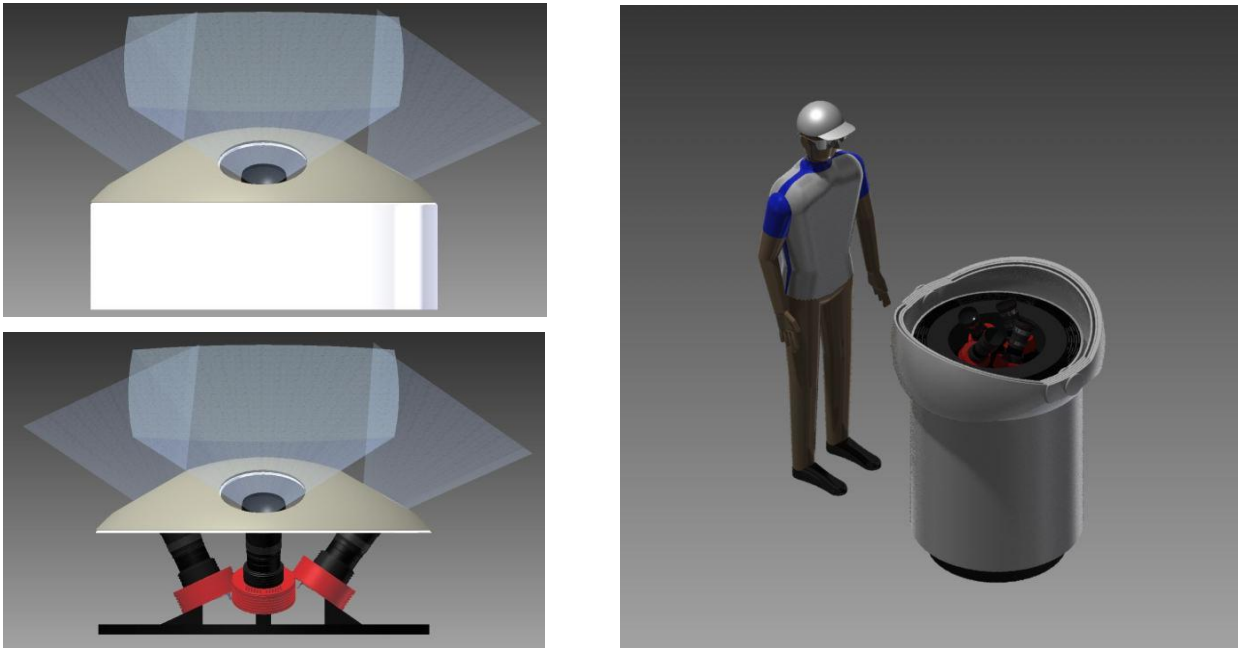

Figure 4: (Left) A MASCARA housing-design option consisting of a pyramid-like structure with 4 sliding covers. (Right) Another design option showing a clam-shell type dome.

## 5. DISCUSSION - COMPARISON TO OTHER INITIATIVES

We plan to roll out the first MASCARA station before the end of 2013, and to deploy the remaining stations in the following two years. Each station will deliver on a clear night light curves for all stars with V=4-8 at airmass<3, at a precision of 1 to 10 millimagnitude per hour. MASCARA could potentially discover up to half a dozen of each class of hot Jupiters, Neptunes, and super-Earths. Several other initiatives, both form the ground and in space, are under planning or development with a similar goal of finding bright transiting planet systems. It is interesting to compare these with MASCARA.

Space missions dedicated to find bright transiting planet systems are proposed both for NASA (TESS[15]) and ESA (PLATO[16]), underlying the importance of this science case. Of course, MASCARA, on face value, cannot compete with either of these missions, most importantly because it will have a much lower instantaneous precision. MASCARA is only sensitive to hot systems (for which many transit events can be averaged), while both PLATO and TESS will be sensitive to planets in much wider orbits. However, MASCARA will be built for only a tiny fraction of the cost of either of these space missions, and will be operational at a significantly shorter time scale.

An alternative concept to detect transiting exoplanets is by using small space telescopes. The Canadian photometric monitoring satellite[17] MOST is playing a pioneering role in this, and discovered the super-Earth 55 Cnc b to transit its naked-eye star, together with NASA's Spitzer Space Telescope[9-10]. However, since only one star at the time can be observed, one can only effectively target known exoplanet systems. The Massachusetts Institute of Technology is leading a study on ExoplanetSat[18] which takes this one step further by utilizing (relatively) low-cost nano-satellites.

Another powerful way to find the transiting planets whose atmospheres can be studied best, is by searching for planets orbiting M-dwarfs. This route is taken by the MEarth team lead by David Charbonneau at Harvard University, who found their first transiting super-Earth[19] Gj1214b. Late M-dwarfs have the great benefit that they are significantly smaller than solar-type stars, making potential transit signals up to two orders of magnitude larger. A similar approach will be taken by the Next Generation Transit Survey[20] (NGTS), which will be sited at ESO-Paranal. This method is highly complementary to MASCARA, since its probes a very different (and fainter) stellar population.

This work is supported by the Dutch Research School for Astronomy (NOVA).

## REFERENCES

[1] Barge, P. et al., "Transiting exoplanets from the CoRoT space mission : 1. CoRoT-Exo-1b: a low-density short-period planet around a G0V star" A&A 482, L17-L20 (2008)
[2] Borucky, W. J., et al., "Kepler Planet-Detection Mission: Introduction and First Results" Science 327, 977-980 (2010)
[3] Leger, A., et al. "Transiting exoplanets from the CoRoT space mission VIII. CoRoT-7b: the first Super-Earth with measured radius", A&A 506, 287-302 (2009)
[4] Bathalia N., et al., "Kepler's first rocky planet: Kepler-10b", ApJ 729, 27 (2011)
[5] Deeg, H., et al. "A transiting giant planet with a temperature between 250 K and 430 K", Nature 464, 384-387 (2010)
[6] Lissauer, J., et al. "A Closely-Packed System of Low-Mass, Low-Density Planets Transiting Kepler-11", Nature 470, 53–58 (2011)
[7] Mayor M. et al., "The HARPS search for southern extra-solar planets. XIII. A planetary system with 3 super-Earths (4.2, 6.9, and 9.2 Me)", A&A 493, 639 (2009)
[8] Howard A. et al., "The Occurrence and Mass Distribution of Close-in Super-Earths, Neptunes, and Jupiters", Science 330, 653 (2010)

[9] Winn J. et al., "A Super-Earth Transiting a Naked-eye Star", ApJ 737, 18 (2011)
[10] Demory B. et al., "Detection of a transit of the super-Earth 55 Cnc e with Warm Spitzer", A&A 533, 114 (2011)
[11] Pollacco, D. L., et al. "The WASP Project and the SuperWASP Cameras", The Publications of the Astronomical Society of the Pacific 118, 1407-1418 (2006)
[12] Bakos, G. Á. et al., "HAT-P-1b: A Large-Radius, Low-Density Exoplanet Transiting One Member of a Stellar Binary", ApJ 656, 552-559 (2007)
[13] Alonso, R, et al. "TrES-1: The Transiting Planet of a Bright K0 V Star", ApJ 613, L153 (2004)
[14] Bouchy F. et al." ELODIE metallicity-biased search for transiting Hot Jupiters: II. A very hot Jupiter transiting the bright K star HD 189733", A&A 444, L15-L19 (2005)
[15] Ricker, G., "The Transiting Exoplanet Survey Satellite (TESS)", AAS 21340301 (2009)
[16] Catala, C., et al. "PLATO : PLAnetary Transits and Oscillations of stars", Journal of Physics: Conference Series, Volume 271, Issue 1, pp. 012084 (2011)
[17] Rowe, J. F. et al. "An Upper Limit on the Albedo of HD 209458b: Direct Imaging Photometry with the MOST Satellite", ApJ 646, 1241 (2006)
[18] Smith M. et al., "ExoplanetSat: detecting transiting exoplanets using a low-cost CubeSat platform", SPIE 7731, 66 (2010)
[19] Charbonneau, D. et al., "A super-Earth transiting a nearby low-mass star", Nature 462, 891 (2009)
[20] http://www.ngtransits.org/